\documentclass[11pt]{article}

\usepackage{amsxtra,amssymb,amsthm,amsmath,latexsym}
\usepackage{graphicx}
\usepackage{epsfig}
\usepackage{afterpage}

\newcommand{\R}{{\mathbb R}}

\newcommand{\dl}{{\delta}}

\newcommand{\bee}{\begin{equation*}}
\newcommand{\eee}{\end{equation*}}
\newcommand{\be}{\begin{equation}}
\newcommand{\ee}{\end{equation}}
\newcommand{\pn}{\par\noindent}
\title{Boundary integral equation for electromagnetic wave scattering 
by a homogeneous body of arbitrary shape}
\author{A G Ramm\\
\small Department of Mathematics\\[-0.8ex]
\small Kansas State University, Manhattan, KS 66506-2602, USA\\[-0.8ex]
\small \texttt{ramm@math.ksu.edu}\\
}

\begin{document}
Intern.Journ. Pure and Appl. Math., 55, N4, (2009). 13-16.
\date{}
\maketitle

\begin{abstract}Boundary integral equation is derived for
the problem of scattering of electromagnetic waves by 3D
homogeneous body of arbitrary shape.
\end{abstract}
\pn{
 }

\section{Introduction}
Let $D\subset \R^3$ be a bounded domain with a smooth connected
boundary $S$. The exterior domain $D'=\R^3\setminus D$ is filled with 
a
homogeneous material with parameters $(\epsilon_0,\sigma=0,\mu_0)$,
$D$ is filled with a material with parameters
$(\epsilon,\sigma,\mu_0)$, $\epsilon,$ $\epsilon_0$, $\sigma$,
$\mu_0$ are constants. Let
$\epsilon'=\left\{
            \begin{array}{ll}
              \epsilon+\frac{i\sigma}{\omega}, & \hbox{in $D$,} \\
              \epsilon_0, & \hbox{in $D'$.}
            \end{array}
          \right.
$ 
Here $\omega$ is the frequency, $\mu_0$ is magnetic parameter,
$\epsilon,\epsilon_0$ are dielectric parameters, $\sigma\geq 0$ is
conductivity.

The governing equations in $\R^3$ are \be\label{e1} \nabla\times
E=i\omega \mu_0H,\quad \nabla\times H=-i\omega \epsilon' E. \ee The
boundary conditions on $S$ are continuity of the tangential
components of $E$ and normal components of $\epsilon'E$ across $S$:
\be\label{e2} [N,E^+]=[N,E^-], \ee and \be\label{e3} N\cdot
(\epsilon' E^+)=N\cdot( \epsilon_0 E^-). \ee Here $N$ is the unit
normal to $S$, pointing into $D'$, $[N,E]$ ($E\cdot N$) is the cross
(dot) product of vectors, and $E^+(E^-)$ is the limiting value of
$E$ as $x\to s\in S$, $x\in D (D')$.

We assume that the incident field $(E_0,H_0)$ satisfies equations
\eqref{e1} in $\R^3$ with $\epsilon'$ replaced by $\epsilon_0$. For 
example
one may take the incident field to be a plane wave: 
$E_0=e_1e^{ikx_3},$ where $e_i\cdot e_j=\dl_{ij}$,
$\{e_j\}_{j=1}^3$ is the standard Euclidean basis,
$\dl_{ij}=\left\{
            \begin{array}{ll}
              1,  & \hbox{if $i=j$,} \\
              0, & \hbox{if $i\neq j$.}
            \end{array}
          \right.
$

If $E$ is found
then 
\be\label{e4} H=(i\omega \mu_0)^{-1}\nabla \times E. 
\ee 
From
\eqref{e1} one gets 
\be\label{e5} \nabla\times\nabla\times
E-\mathcal{K}^2E=0\qquad in \ \R^3, \ee where\be\label{e6}
\mathcal{K}^2=\left\{
                                                 \begin{array}{ll}
                                                   K^2\ in\ D, & \hbox{$K^2=\omega^2\epsilon'\mu_0$,} \\
                                                   k^2\ in\ D', & \hbox{$k^2=\omega^2\epsilon_0\mu_0$.}
                                                 \end{array}
                                               \right.
 \ee The scattering problem consists of finding the solution of equations \eqref{e5}, \eqref{e2}, \eqref{e3}, such that
\be\label{e7} E=E_0+V,\ee \be\label{e8}
V_r-ikV=o\left(\frac{1}{r}\right),\quad r:=|x|\to \infty.\ee Assumption
\eqref{e8} means that $V$ satisfies the radiation condition.

\section{Derivation of the boundary integral equations}
Let us
look for the solution to \eqref{e5}, \eqref{e2}, \eqref{e3},
\eqref{e7}, \eqref{e8}, of the form: \be\label{e9} E=\left\{
                                                       \begin{array}{ll}
                                                         \nabla\times \int_SG(x,t)J(t)dt, & \hbox{$x\in D$} \\
                                                         \nabla\times \int_Sg(x,t)j(t)dt+E_0(x), & \hbox{$x\in D'$,}
                                                       \end{array}
                                                     \right.
 \ee where
\be\label{e10} g(x,y)=\frac{e^{ik|x-y|}}{4\pi|x-y|},\quad
G(x,y)=\frac{e^{iK|x-y|}}{4\pi|x-y|}. \ee 
The currents $j$ and $J$ are vector
fields tangential to $S$. 

Thus, there are four scalar unknowns: the two
unknown vector fields $j$ and $J$, tangential to $S$. 

For any $j$
and $J$ vector $E$ solves equation \eqref{e5} in $D$ and in $D'$ and
satisfies conditions \eqref{e7} and \eqref{e8} because $g$ satisfies
the radiation condition \eqref{e8}. Thus, \eqref{e9} is the solution
to the scattering problem if $j$ and $J$ can be chosen so that the
boundary conditions \eqref{e2}, \eqref{e3} are satisfied.

Condition \eqref{e2} is a vector equation, which is equivalent to
three scalar equations, and equation \eqref{e3} is a scalar
equation. Therefore equations \eqref{e2} and \eqref{e3} together are
equivalent to four scalar equations for four unknown scalar functions, the
coordinates of the tangential to $S$ vector fields  $j$ and $J$.

Equation \eqref{e2} can be written as a Fredholm-type integral
equation. We have

\be\label{e11} \int_S[N_x,[\nabla_xG(x,t),J]]|_{x\to s,\ x\in
D}=-\frac{A^+J+J}{2}+\int_S\nabla_sG(s,t)N_s\cdot J(t)dt,\ee where
the known formula for the limiting value of the normal derivative of
the single-layer potential was used (see, e.g., \cite{M}).
 
Since $J(s)\cdot N_s=0$
(because $J(s)$ is a tangential vector field) and $J(s)$ is assumed
Lipschitz, the last integral in \eqref{e11} converges absolutely.
Since we assume the surface sufficiently smooth, e.g., $S\in C^{1,a}$
$a>0$, and the incident field is smooth, the currents $j$ and $J$
are as smooth as the data, in particular, they are Lipschitz.
The class of surfaces, satisfying the condition  $S\in C^{1,a}$,
consists of surfaces whose graph in local coordinates is 
differentiable and its derivative satisfies the H\"older condition with the exponent 
$a\in (0,1]$.
 
The operator $A^+$ in \eqref{e11} is defined as \be\label{e12}
A^+J=\int_S\frac{\partial G(s,t)}{\partial N_s}J(t)dt. \ee Thus,
equation \eqref{e2} can be written as: \be\label{e13}\begin{split}
-\frac{A^+J+J}{2}+\int_S\nabla_sG(s,t)N_s\cdot
J(t)dt&=-\frac{A^-j+j}{2}\\
&+\int_S\nabla_sg(s,t)N_s\cdot j(t)dt+[N,E_0],
\end{split}
\ee 
where
\be\label{e14}A^-j=\int_S\frac{\partial g(s,t)}{\partial
N_s}j(t)dt.\ee 
Equation \eqref{e13} is of Fredholm type: the
integral operators in \eqref{e13} are compact in $C(S)$ and in
$L^2(S)$. Equation \eqref{e3} yields: \be\label{e15}
N_s\cdot\int_S[\nabla_xG(x,t),J(t)]|_{x\to s,\ x\in D}dt=N_s\cdot
\int_S[\nabla_xg(x,t),j(t)]|_{x\to s,\ x\in D'}dt+N_s\cdot E_0. \ee
Equation \eqref{e15} is singular. 

{\bf Claim:} {\it The integrals in \eqref{e15} exist
as Cauchy principal values.}

Let us verify this claim. One has
\be\label{e16}\begin{split}
N_s\cdot [\nabla_x G(x,t),J(t)]_{x\to D,\ x\in D}&=N_s\cdot
[\frac{e^{iKr_{st}}}{4\pi
r_{st}}(iK-\frac{1}{r_{st}})r^0_{ts},J(s)]\\
&+O\left(\frac{1}{r_{st}}\right),\quad
r^0_{ts}:=\frac{\vec{r}_{ts}}{r_{st}},\ r_{st}=|\vec{r}_{ts}|,
\end{split}\ee
because $|J(s)-J(t)|\leq c|s-t|.$ 

The singular term in \eqref{e16} is
\be\label{e17} \frac{1}{4\pi r^2_{st}} N_s\cdot
[r^0_{st},J(s)]=\frac{|J(s)|}{4\pi r^2_{st}}\sin\theta, \ee where
$\theta=\theta(s,t)$ is the angle between the $x$-axis and the
vector $\vec{r}_{st}$. We choose the $x$- axis in the plane tangential to $S$
at the point $s$ so that it is directed along the vector $J(s)$. 

Since\bee
\int_0^\pi \sin\theta d\phi=0,\quad t=e^{i\phi},\ 0\leq \phi<2\pi,
\eee 
the {\bf Claim} follows from Theorem 1.1 on p. 221 in \cite{MP}. This
theorem says that a singular integral
$\int_{\R^m}\frac{f(x,\theta)}{|x-y|^m}u(y)dy$ exists as a Cauchy
principal value if 
$$\int_{S^{m-1}}f(x,\theta)ds=0,\quad
\theta=\frac{y-x}{|y-x|}.$$ 

Numerical methods for solving Fredholm
equations and singular integral equations are well developed
(\cite{MP}). They are not discussed here.


\begin{thebibliography}{00}
\bibitem{MP} S. Mikhlin, S. Pr\"ossdorf, Singular integra; operators, 
Springer-Verlag, Berlin, 1986.
\bibitem{M} C. M\"uller, Foundations of the mathematical theory of
electromagnetic waves, Springer-Verlag, Berlin, 1969.
\end{thebibliography}
\end{document}